\documentclass[aps,prb,reprint,showpacs,longbibliography,superscriptadress]{revtex4-1}
\usepackage{amsmath,amssymb,bm}
\usepackage{graphicx}
\usepackage{epstopdf}
\usepackage{latexsym}
\usepackage{color}
\usepackage{hyperref}

\newcommand{\sulf}{{Sul-Cu$_2$Cl$_4$}}
\newcommand{\sulfful}{Cu$_2$Cl$_{4}\cdot$H$_8$C$_4$SO$_2$}
\newcommand{\aver}[1]{\left\langle #1 \right\rangle}

\bibpunct{[}{]}{,}{n}{}{}

\begin{document}

\title{Giant dielectric nonlinearities at a magnetic Bose--Einstein condensation}

\author{K. Yu. Povarov}
    \email{povarovk@phys.ethz.ch}
   \affiliation{Neutron Scattering and Magnetism, Laboratory for Solid State Physics, ETH Z\"{u}rich, Switzerland}
   \homepage{http://www.neutron.ethz.ch/}

\author{A. Reichert}
    \affiliation{Neutron Scattering and Magnetism, Laboratory for Solid State Physics, ETH Z\"{u}rich, Switzerland}

\author{E. Wulf}
    \affiliation{Neutron Scattering and Magnetism, Laboratory for Solid State Physics, ETH Z\"{u}rich, Switzerland}

\author{A. Zheludev}
    \email{zhelud@ethz.ch}
    \affiliation{Neutron Scattering and Magnetism, Laboratory for Solid State Physics, ETH Z\"{u}rich, Switzerland}

\date{\today}

\begin{abstract}
We experimentally investigate the dielectric response of the
low-dimensional gapped quantum magnet \sulfful\ near a magnetic
field-induced  quantum critical point, which  separates the
quantum-disordered and helimagnetic ground states. The observed
magnetocapacitive effect originates from an \emph{improper}
ferroelectric nature of the transition, which itself is perhaps one
of the best known realizations of Bose--Einstein condensation of
magnons. Despite that, we find that the magnetocapacitive effect
associated with the transition exhibits huge and very unusual
anharmonicities.
\end{abstract}



\maketitle

The phenomenon of magnetic Bose--Einstein condensation (BEC) is one
of the most fascinating analogies between the properties of quantum
magnets and bosonic
systems~\cite{Giamarchi_NatPhys_2008_BECreview,Zapf_RMP_2014_BECreview}.
It corresponds to spontaneous long-range ordering induced in gapped
quantum paramagnets by external magnetic fields. Since conventional
BEC is a spontaneous breaking of $U(1)$ gauge symmetry, a crucial
requirement for the magnetic analog is the presence of isomorphous
$SO(2)$ axial spin rotation symmetry in the disordered phase.
Unfortunately,  symmetry-breaking anisotropic interactions that are
always present in real crystalline materials preclude the
applicability of the BEC concept close to the quantum critical
point. The only exception would be the case of incommensurate
helimagnetic order. Here, the required $SO(2)$ symmetry is robust,
as it represents a sliding of the magnetic spiral with respect to
the crystal lattice, made ``frictionless'' by its
incommensurability. There is, however, a potential complication.
Helimagnets are likely to be
multiferroic~\cite{CheongMosotovoy_NatMat_2007_NatureReview}. In a
spiral structure, the lack  of inversion symmetry between the spins
gives rise to electric
polarization~\cite{Katsura_PRL_2005_MFmicro,Mostovoy_PRL_2006_MFmacro,SergienkoDagotto_PRL_2007_DualMF}.
The intrinsic coupling between magnetic and electric degrees of
freedom is a complication that can potentially modify the very
nature of the quantum critical point (QCP). A few recent studies
looked at rare cases in which multiferroicity occurs at quantum
phase transitions
(QPTs)~\cite{Kenzelmann_PRL_2007_TriangularFerro,Abe_PRB_2009_TbMnO3Flop,Kim_NatComm_2014_SaturationSusceptibility},
but not much work was done to study the criticality of such
transitions. Moreover, the very important case of QCPs connecting
magnetically ordered and quantum-disordered phases remains largely
unexplored in the context of multiferroic physics. The two questions
to be addressed are as follows: (i) Can this type of transition be
described as BEC and (ii) how do the dielectric properties behave at
and beyond the QCP?

On the experimental side, some progress has occurred very recently
with the discovery of the $S=1/2$ quasi-one-dimensional quantum
antiferromagnet \sulfful\ (also known as
\sulf)~\cite{Fujisawa_ProgThPhys_2005_SulfHC,GarleaZheludev_PRL_2008_SulfZF,GarleaZheludev_PRB_2009_SulfOrdering,ZheludevGarlea_PRB_2009_SulfExcitations}.
This material features a magnetic-field-induced QPT from a
quantum-disordered to a helimagnetic state. However, previous
studies have portrayed this transition as being more complex than a
simple magnetic BEC. Dielectric spectroscopy
experiments~\cite{Schrettle_PRB_2013_SulMF} pointed to the
possibility of a dielectric susceptibility divergence at $T=0$. At
the same time, the shape of the phase boundary and the order
parameter exponent were found to be inconsistent with the BEC
paradigm~\cite{Fujisawa_ProgThPhys_2005_SulfHC,GarleaZheludev_PRB_2009_SulfOrdering}.
This indicated the possibility of an unconventional order parameter
involving both spin and charge degrees of freedom. In the present
Rapid Communication, we show that this is \emph{not} the case, and
that electric polarization in \sulf\ is \emph{not} a true order
parameter of the transition. We show that the QCP is a ``protected''
BEC of magnons, and is arguably the only clean realization of such
among all known materials. While electric polarization plays a
dependent role, its behavior is far from trivial. Near the magnetic
QCP in \sulf, we find spectacular anomalies in the \emph{nonlinear}
dielectric response.

\begin{figure}[!t]
\includegraphics[width=0.45\textwidth]{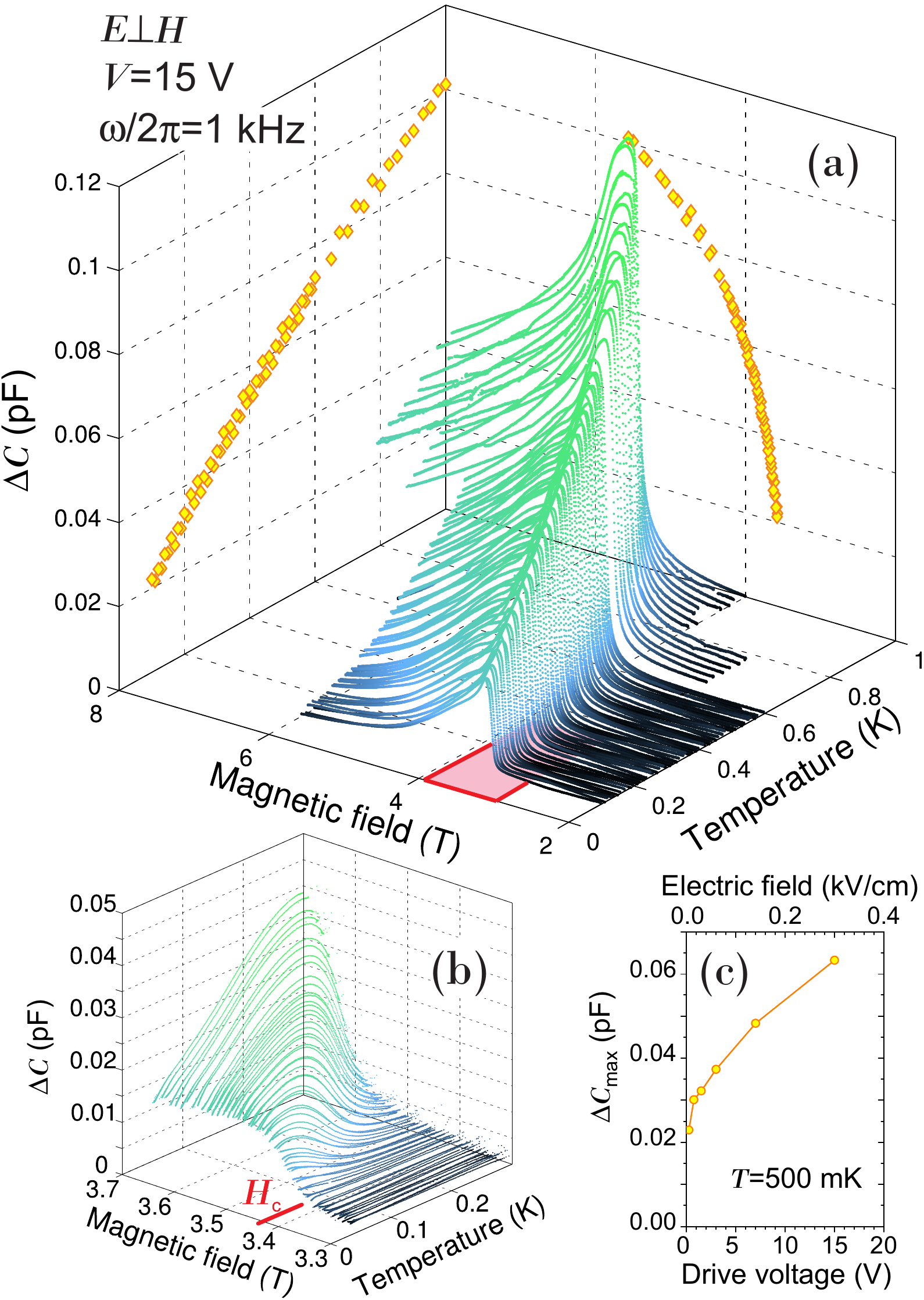}
\caption{(Color online) Magnetocapacitive effect in \sulf. The
anomalous contribution to the sample capacitance $\Delta C$ is shown
as a function of temperature and magnetic field. (a) Field scans at
constant temperature. The peak positions and amplitudes are
projected onto the $\Delta C-H$ and $\Delta C-T$ planes (yellow
diamonds). The highlighted $H-T$ region in close vicinity to the QCP
is the domain of temperature scans. (b) Temperature scans. The step
in field is $0.002$~T. The extrapolated value of the
zero-temperature critical field is marked. (c) Measured dependence
of the amplitude of the dielectric anomaly on the probing voltage at
$T=500$~mK.}\label{FIG:sul_T}
\end{figure}

\begin{figure}[]
  \includegraphics[width=0.5\textwidth]{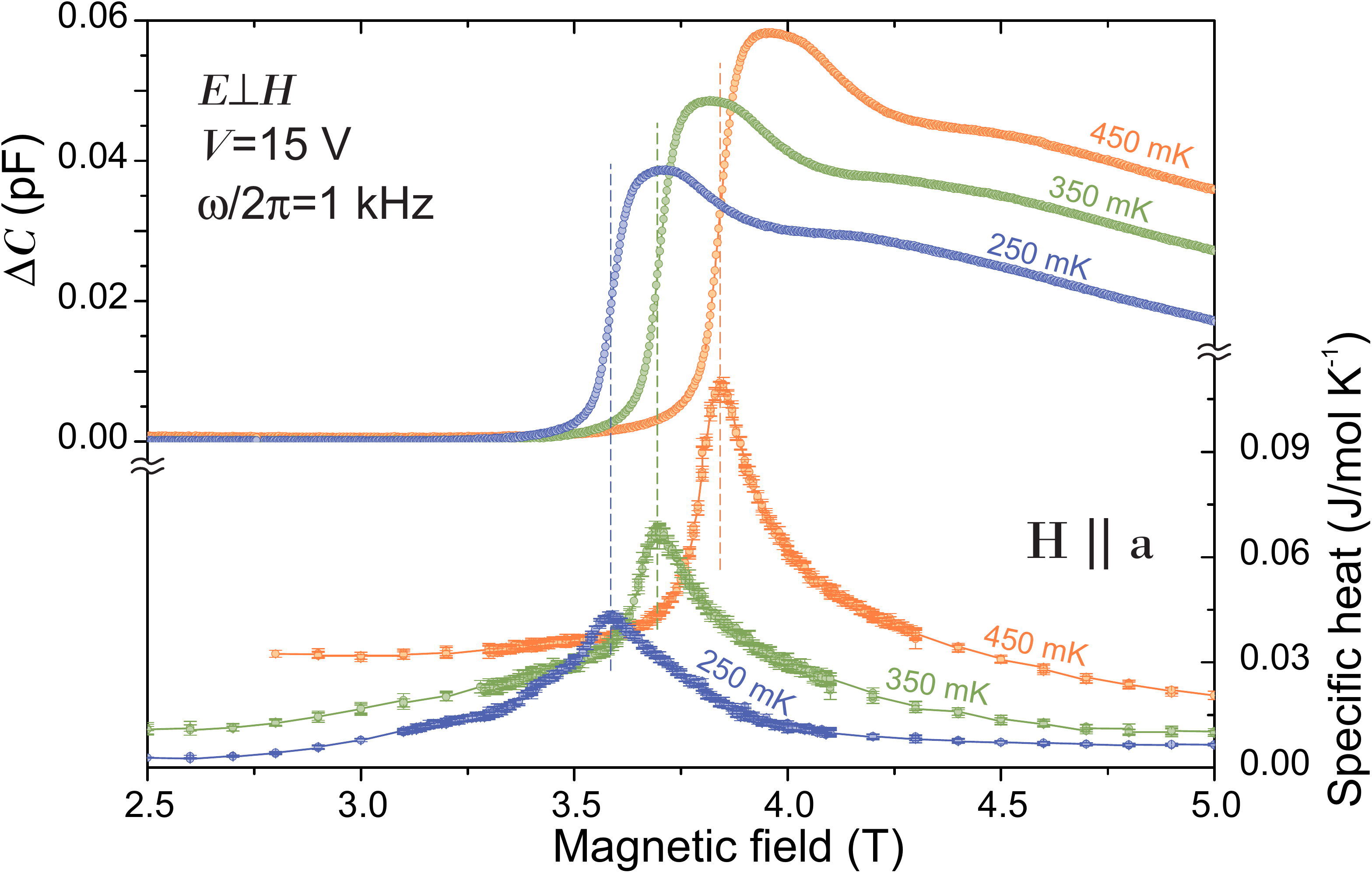}\\
  \caption{(Color online) A few representative field scans
  of a sample capacitance (top), compared to the specific heat data (bottom).
  For both measurements, $\mathbf{H}\parallel\mathbf{a}$.}\label{FIG:compare}
\end{figure}

\begin{figure}[!t]
  \includegraphics[width=0.5\textwidth]{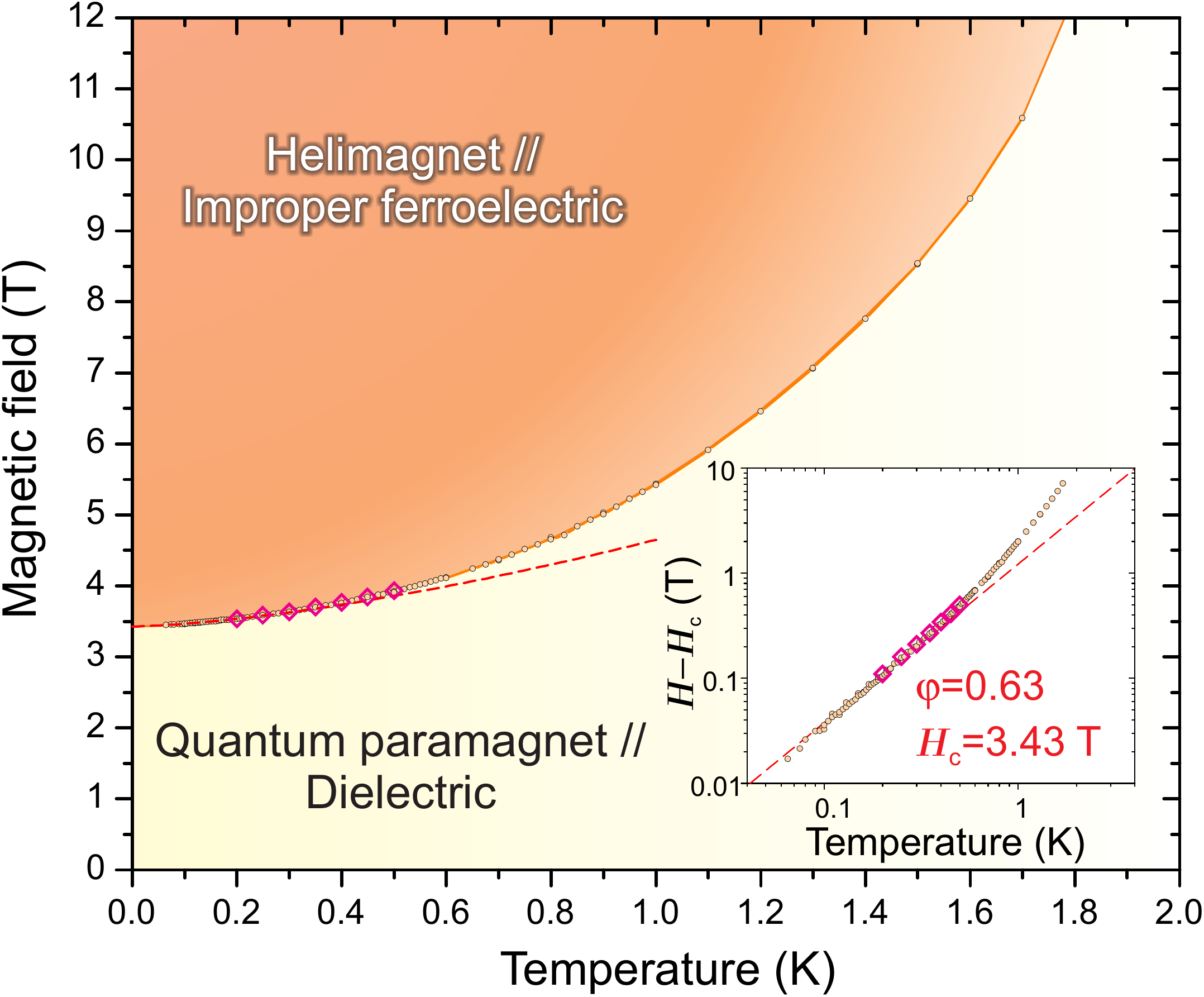}\\
  \caption{(Color online) Phase diagram of \sulf\ for $\mathbf{H}\parallel\mathbf{a}$.
  Small circles are the transition points determined from the magnetocapacitive effect and diamonds
  are the transition points found from the specific heat $\lambda$-anomaly.
  The solid line is a guide for the eye. The dashed line is a fit to Eq.~(\ref{EQ:crossover}) in
  the range $T<300$~mK. The inset shows a logarithmic plot of the measured phase boundary.}\label{FIG:diagram}
\end{figure}

\sulf\ belongs to a family of insulating metaloorganic Heisenberg
spin systems. Its magnetic properties are due to $S=1/2$ Cu$^{2+}$
ions. The spins are antiferromagnetically coupled into
one-dimensional structures, which can be described as highly
frustrated four-leg spin tubes, running along the $c$ axis of the
triclinic structure (see
Refs.~\cite{GarleaZheludev_PRL_2008_SulfZF,GarleaZheludev_PRB_2009_SulfOrdering}
for more details). This layout, with an even number of ``legs'', is
responsible for the nonmagnetic quantum-disordered ground state, and
a gap $\Delta\simeq0.52$~meV in the spin excitation spectrum. Due to
a geometric frustration of the exchange interactions, the dispersion
minimum for the triplet of lowest-energy $S=1$ excitations occurs at
an incommensurate wave vector. As a result, in an external magnetic
field $H_{c}=\Delta/g\mu_{\text{B}}\sim 3.7$~T for
$\mathbf{H}||\mathbf{b}$, \sulf\ undergoes a soft-mode ordering
transition with a propagation vector $\mathbf{Q}=(-0.22,0,0.48)$
~\cite{GarleaZheludev_PRB_2009_SulfOrdering,ZheludevGarlea_PRB_2009_SulfExcitations}.
The structure of the ordered high-field phase has been resolved by
neutron diffraction~\cite{GarleaZheludev_PRB_2009_SulfOrdering}. The
spin components transverse to $\mathbf{H}$ form a spiral such that
$\aver{\mathbf{S}_{\perp}(\mathbf{r})}=\mathbf{S}_{1}\cos(\mathbf{Qr})+\mathbf{S}_{2}\sin(\mathbf{Qr})$
with $\mathbf{S}_{1}\perp\mathbf{S}_{2}$. This spiral arrangement
lacks an inversion symmetry and hence induces an electric
polarization
$\mathbf{P}\propto\boldsymbol{\left[}\left[\mathbf{S}_{1}\times\mathbf{S}_{2}\right]\times\mathbf{Q}\boldsymbol{\right]}$
~\cite{Katsura_PRL_2005_MFmicro,Mostovoy_PRL_2006_MFmacro}. To date,
this polarization has not been directly measured, and is likely to
be extremely small. Instead,  previous studies of
Schrettle~\textit{et al.}~\cite{Schrettle_PRB_2013_SulMF} probed the
corresponding dielectric permittivity $\varepsilon$. The apparent
divergence of this quantity at the phase transition was taken as a
sign of the critical fluctuations of $P$, which would imply that the
transition is different from simple ordering of localized spins.

The main limitation of that study was in that it only probed
temperatures above $1$~K, never closely approaching the QCP at
$T=0$. In the present work we have overcome this technical
difficulty by combining a capacitance bridge setup similar to that
used in Ref.~\cite{Schrettle_PRB_2013_SulMF} with a $^3$He-$^4$He
dilution refrigerator~\cite{SuppMat}. What is in fact measured in
our experiments is the capacitance of a plate capacitor with a
\sulf\ sample in between the plates. At low temperatures in the
absence of magnetic field it is dominated by a constant background
$C_{0}\simeq3.5$~pF due to the ``normal'' dielectric permittivity of
\sulf, $\varepsilon\simeq3$. The quantity of interest is the
additional capacitance $\Delta C(H,T)$, which is both temperature
and magnetic field dependent near the QCP. It is plotted in
Figs.~\ref{FIG:sul_T}(a) and \ref{FIG:sul_T}(b) and represents the
observed magnetocapacitive effect, with a maximal detected
$\Delta\varepsilon/\varepsilon\sim3$\%.

The prominent capacitance peak seen at \emph{high} temperatures is
what Ref.~\cite{Schrettle_PRB_2013_SulMF} must have taken for a
divergent $\varepsilon$. In fact, at low temperatures, as the QCP is
approached, the anomaly weakens, and transforms into a small and
rounded step at $T\rightarrow 0$. The peak amplitude of $\Delta C$
decreases with the temperature almost linearly, rather than
diverging at the QCP. A comparison between the capacitance peaks and
$\lambda$-anomalies in specific heat for
$\mathbf{H}\parallel\mathbf{a}$ (Fig.~\ref{FIG:compare}) shows that
the transition point corresponds to the inflection point of the
$\Delta C(H)$ steep slope. Note that critical fluctuations are
prominent in specific heat well below the transition field, while
the dielectric contribution immediately disappears in the disordered
phase. Moreover, as shown in Fig.~\ref{FIG:sul_T}(b), there is no
detectable change in the dielectric permittivity outside of the
ordered phase \emph{even along the critical trajectory}
$H=H_{c}=3.43\pm0.01$~T~\cite{SuppMat}. We conclude that below
$H_{c}$ there are no critical fluctuations of polarization. The
anomalous contribution to dielectric susceptibility is confined to
the ordered phase and is triggered by the spontaneous magnetic
order. The situation is akin to conventional thermal phase
transitions in improper ferroelectrics, where $P$ appears not in a
spontaneous way, but as a result of coupling to some other parameter
which actually undergoes
criticality~\cite{Dvorak_Ferr_1974_improperReview}. Thus, the QCP in
\sulf\ is related to the magnetic degrees of freedom alone, while
the field-induced phase is an {\it improper} ferroelectric.

Although ferroelectricity does not drive the QPT in \sulf, the very
precise measurements of $\varepsilon$ enable us to extract the
critical field with very high accuracy for this compound, and
thereby elucidate the nature of the magnetic transition
(Fig.~\ref{FIG:diagram}). The parameter most relevant to the QCP is
the so-called crossover exponent $\varphi$, which defines the phase
boundary at $T\rightarrow 0$:
\begin{equation}
\label{EQ:crossover}
    H-H_{c}\propto T^{1/\varphi}.
\end{equation}
A  shrinking fit window analysis~\cite{SuppMat} of our data reveals
that a true power-law behavior extends only up to $T\sim300$~mK, a
range not probed by previous specific heat
studies~\cite{Fujisawa_ProgThPhys_2005_SulfHC,WulfMuhlbauer_PRB_2011_SulDisordered}.
However, as can be seen from the Fig.~\ref{FIG:diagram} inset, our
dielectric data are dense and accurate enough to yield a very
reliable estimate $\varphi=0.63\pm0.03$ for that fitting range. The
anomalous result $\varphi\simeq0.34$ previously obtained by neutron
scattering~\cite{GarleaZheludev_PRB_2009_SulfOrdering} is likely due
to the fact that of the five data points obtained in this study, all
except one did lie outside of the true power-law range. Our present
result $\varphi\simeq0.63$ is fully consistent with $\varphi=2/3$
expected for a magnetic BEC
transition~\cite{Giamarchi_NatPhys_2008_BECreview,Zapf_RMP_2014_BECreview}.

This finding is significant. Indeed, despite numerous claims to the
contrary, the field-induced transition in most gapped quantum
magnets is, strictly speaking, \emph{not} in the BEC universality
class. Instead, due to the magnetic anisotropy which breaks the
prerequisite $SO(2)$ symmetry of the Hamiltonian in materials such
as TlCuCl$_3$ and IPA-CuCl$_3$,  it is actually of the Ising
type~\cite{DellAmore_PRB_2009_BECbreakdown} with an energy gap in
the magnetically ordered
state~\cite{Glazkov_PRB_2004_TlCuCl3gap,Sirker_EurPLett_2004_SOCBEC}.
Even in the one known tetragonal compound
DTN~\cite{Zapf_PRL_2006_BECinDTN,YinXia_PRL_2008_DTNcritical} the
transition is likely discontinuous due to magnetoelastic coupling
and a spontaneous lattice
distortion~\cite{DellAmore_PRB_2009_BECbreakdown}, showing critical
exponents inconsistent with
BEC~\cite{WulfHuvonen_PRB_2015_DTNIntrinsicBroadening}. In contrast
to all these commensurately ordering materials, the $SO(2)\equiv
U(1)$ symmetry in \sulf\ is \emph{protected by its
incommensurability}. Indeed, the phase of the incommensurate spiral
structure is decoupled from any magnetic anisotropy terms that are
commensurate, and the ordered state necessarily has a gapless
``sliding mode''~\cite{BruceCowley_JPC_1978_Phason}.

Even though in \sulf\ the charge degrees of freedom do not drive the
QCP, the dielectric properties here are quite unusual. A careful
examination of the observed magnetocapacitive effect reveals its
amazingly nonlinear nature. Just varying the probing voltage leads
to a drastic change of the amplitude of the dielectric anomaly,
without affecting the onset of the transition or the background
$C_{0}$. An example of such dependence is shown in
Fig.~\ref{FIG:sul_T}(c). The peak magnitude $\Delta C_{\text{max}}$
at a fixed temperature monotonically decreases with a decrease of
the voltage. Note that the electrical fields producing this
nonlinearity are very modest [upper scale in
Fig.~\ref{FIG:sul_T}(c)].

\begin{figure}[!t]
\begin{center}
        \includegraphics[width=0.5\textwidth]{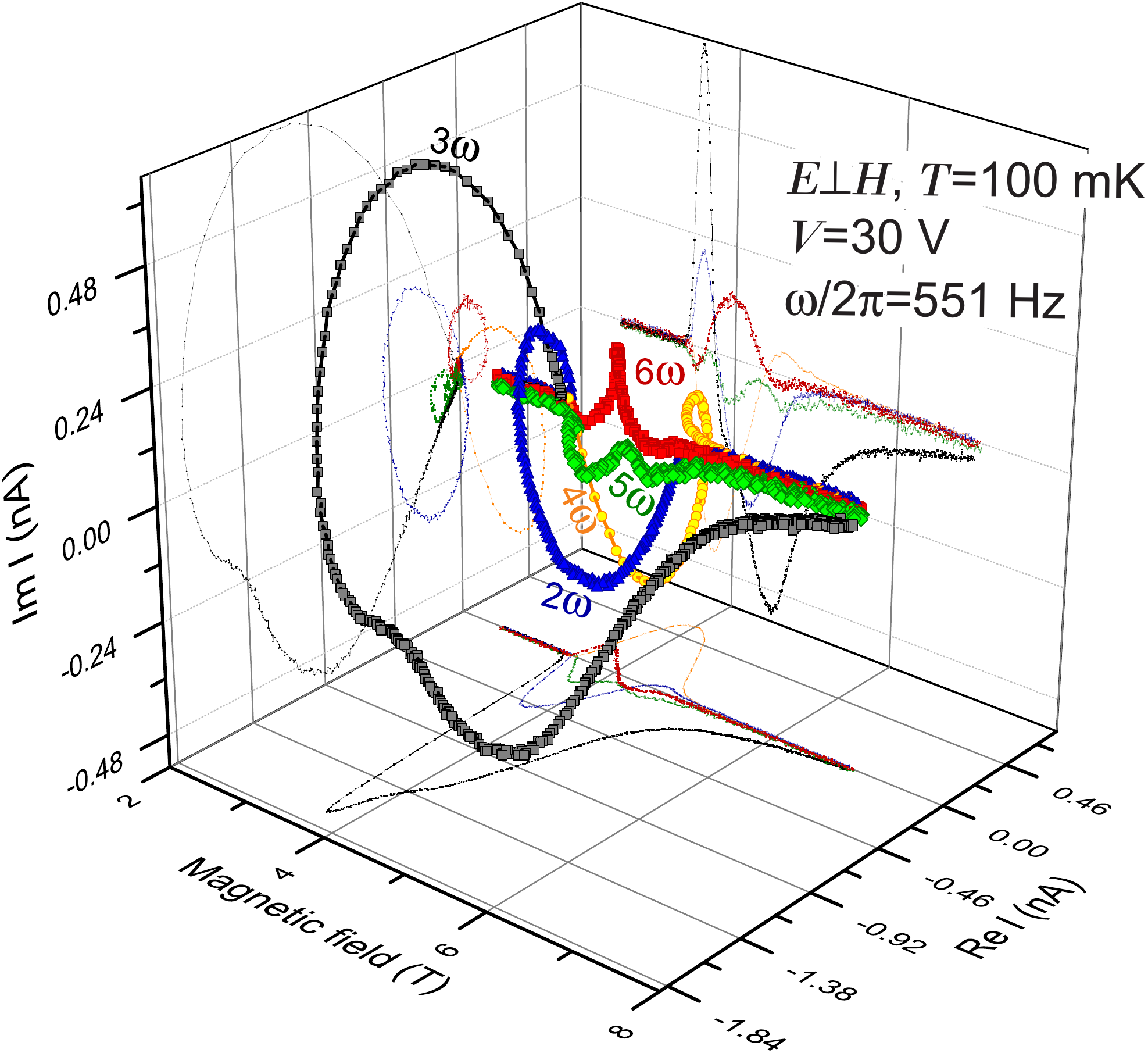}
        \caption{(Color online) Phase transition in \sulf\ as seen by nonlinear dielectric spectroscopy. Current harmonics $I_{n}(H)$  ($n$ from $2$ to $6$) induced by an applied sine voltage are plotted as a functions of magnetic field at $T=100$~mK. Projections of $I_{n}(H)$ onto Re-$H$, Im-$H$ and Re-Im planes are also shown.
        Drive voltage has amplitude $V=30$~V and frequency $\omega/2\pi=551$~Hz.}\label{FIG:Harmspirals}
\end{center}
\end{figure}

\begin{figure}[!t]
\begin{center}
        \includegraphics[width=0.5\textwidth]{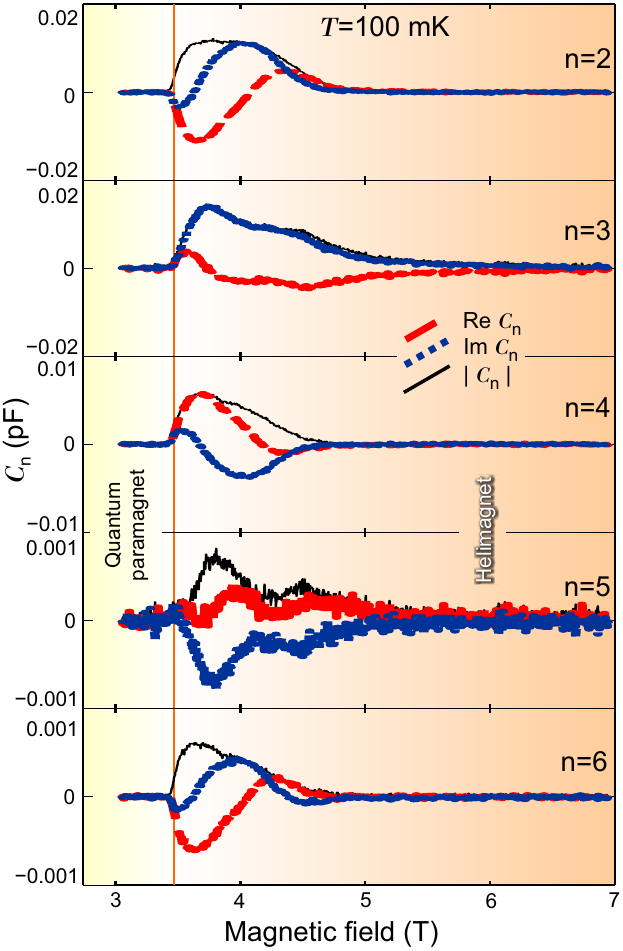}
        \caption{(Color online) Nonlinear contributions to the magnetocapacitive effect in \sulf\ at $T=100$~mK $C_{n}$ (see text) as a functions of magnetic field.
        Thick dashed red and dotted blue lines are real and imaginary parts
        of $C_{n}$
       and thin black line is the absolute value. Vertical line marks the phase transition field.
        }\label{FIG:Hyperboth}
\end{center}
\end{figure}

The best indicators of nonlinear behavior are the higher-order
harmonics in ac measurements. In our case of \sulf, we tracked the
complex harmonics of the displacement current, induced by the
applied ac voltage~\cite{SuppMat}. A representative data set (note
the low $T=100$~mK) is shown in Fig.~\ref{FIG:Harmspirals}. Below
the critical field the nonlinear response is zero. However, inside
the ordered helimagnetic phase all current harmonics with
frequencies up to $6\omega$ show a complicated behavior in the
complex plane as a function of applied magnetic field. At still
higher fields, they gradually decrease and are eventually
suppressed, as is the linear magnetocapacitive effect. Since the
displacement charge flow is related to the change of sample
polarization $\partial P/\partial t$, the appearance of current
harmonics under a periodic voltage $V\sin(\omega t)$ is a direct
consequence of the polarization nonlinearity:

\begin{equation}\label{EQ:nonlinP}
    P(E)=P_{0}+\varepsilon_{0}\left(\chi_{1}E+\chi_{2}E^{2}+\chi_{3}E^{3}+...\right).
\end{equation}

In this expansion $\chi_{1}=\varepsilon-1$ is the conventional
first-order susceptibility, while
$\chi_{n}=\frac{1}{\varepsilon_{0}n!}\frac{\partial^{n}P}{\partial
E^{n}}$ are the nonlinear hypersusceptibilities. The algebra
required to restore the hypersusceptibilities $\chi_{n}$ from
current harmonics $I_{n}$ is summarized in
Ref.~\cite{Miga_RevSciIns_2007_DielectricSpectrometer}. To allow a
{\it quantitative} comparison between different
hypersusceptibilities (which are of different physical dimensions),
we present them in the form of anharmonic capacitance contributions
$C_{n}=\varepsilon_{0}\chi_{n}\frac{S}{d}E^{n-1}$~\cite{SuppMat}.
The resulting $C_{n}(H)$ for the representative $T=100$~mK case are
plotted in Fig.~\ref{FIG:Hyperboth}. By comparing this plot to
Fig.~\ref{FIG:sul_T} one can immediately see that some of the
anharmonic contributions have \emph{the same} order of magnitude, as
the linear magnetocapacitance term. Already at a very moderate
electric field $E\sim0.3$~kV/cm, the nonlinear response is at least
as important as the linear one. Such a giant dielectric nonlinearity
at a quantum phase transition of a nonelectric nature is an
interesting phenomenon.

 A transition-related nonlinear electric response is well
documented for proper ferroelectrics and
relaxors~\cite{Dec_Ferr_2011_DielectricNonlinReview}, but not for
\emph{improper} ferroelectrics. Previous studies of other Cu$^{2+}$
based $S=1/2$ improper helimagnetic ferroelectrics, such as
CuCl$_{2}$~\cite{Seki_PRB_2010_CuCl2Multiferr},
LiCu$_2$O$_2$~\cite{Park_PRL_2007_LiCu2O2Multiferr}, and
LiCuVO$_4$~\cite{Mourigal_PRB_2011_LiCuVO4spincurrent,Ruff_JPCM_2014_LiCuVO4DomainFlops},
have not encountered $P(E)$ nonlinearity at the transition point. A
small nonlinearity was typically found deep in the ordered phase as
a consequence of a well-developed $P(E)$ hysteresis curve. The
direction of $P$ in individual magnetic domains is coupled to the
spiral chirality, which can be switched by sufficiently strong $E$,
producing the hysteresis. In \sulf, a small history dependence of
magnetocapacitance above $H_{c}$ and suppression of the effect by
bias field~\cite{SuppMat} indicate that domains also play an
important role. The observed nonlinearity may in principle originate
from the extreme sensitivity of the domains to $E$ close to the
magnetic order breakdown. Since even tiny amounts of impurities are
known to have a huge effect on the phase transition in
\sulf~\cite{WulfMuhlbauer_PRB_2011_SulDisordered}, in the future it
will be very interesting to investigate their influence on the
dielectric response and domain mobility.

In summary, the field-induced QCP in \sulf\ is a purely magnetic
one, and appears to be one of the best realizations of magnetic BEC.
The material is thus an ``improper quantum field-induced
ferroelectric.'' Its unusual dielectric response is confined to the
magnetically ordered phase and is hugely nonlinear, even in very
modest drive fields.

We would like to extend our thanks to Prof. B. Batlogg for
generously sharing some key equipment for the present experiments,
and to Prof. M. Mostovoy for illuminating discussions. Special
thanks go out to Dr. S. Gvasaliya for assistance with the instrument
development. This work was supported by the Swiss National Science
Foundation, Division 2.

\begin{center}
\textbf{SUPPLEMENTAL MATERIAL}
\end{center}
%
%
%
%

\section{Experimental details}

The sample for the magnetocapacitive measurements was a co-aligned
mosaic of $d\sim0.5$~mm-thin single crystals of \sulf\ was
sandwiched between the copper plates of a capacitor, that were
parallel to the crystallographic $(ac)$ plane. This assembly was
placed at the sample stage of Quantum Design Physical Properties
Measurement System (PPMS) $^3$He-$^4$He Dilution Refrigerator
insert. The magnetic field $\mathbf{H}$ was applied parallel to the
$\mathbf{a}$ axis. It defines the spiral plane, the expected
polarization being along $\mathbf{b^{\ast}}\parallel\mathbf{E}$.
Special care was taken to prevent any exposure of the sample to the
atmosphere to avoid deterioration. Measurements were done using an
AH2550A capacitance bridge. The capacitor plate area was $A =
8\times 8$~mm$^2$. Due to the difficulty of precisely  measuring the
filling factor of the \sulf\ ``effective capacitor'', we prefer to
report the results as a change in sample capacitance
$C=\frac{A}{d}\varepsilon_{0}\varepsilon$ rather than a change in
dielectric permittivity. The values of $E$ and $\varepsilon$
mentioned in the text should be considered as estimates.

Nonlinear dielectric spectroscopy was performed on the very same
sample assembly during the same experimental run. The source voltage
was applied to one of the capacitor plates, and SR7370 lock-in
amplifier was used to read the displacement current harmonics. We
use the actual value of $V=30$~V to estimate the
hypersusceptibilities $\chi_{n}$, but smaller voltage $V=15$~V to
present them as nonlinear capacitances $C_{n}$ for consistent
comparison with the bridge data.

Specific heat was measured on $m\simeq0.1$~mg \sulf\ sample with the
help of adiabatic calorimetry option for PPMS $^3$He-$^4$He Dilution
Refrigerator insert. The contribution of Apiezon N grease and silver
sample holder was measured in a separate run and subtracted. The
measurements extend down to $200$~mK; below this temperature the
data becomes progressively plagued by huge contributions from
nuclear spin specific heat.

\section{Phase diagram}

\begin{figure}[b]
\begin{center}
        \includegraphics[width=0.4\textwidth]{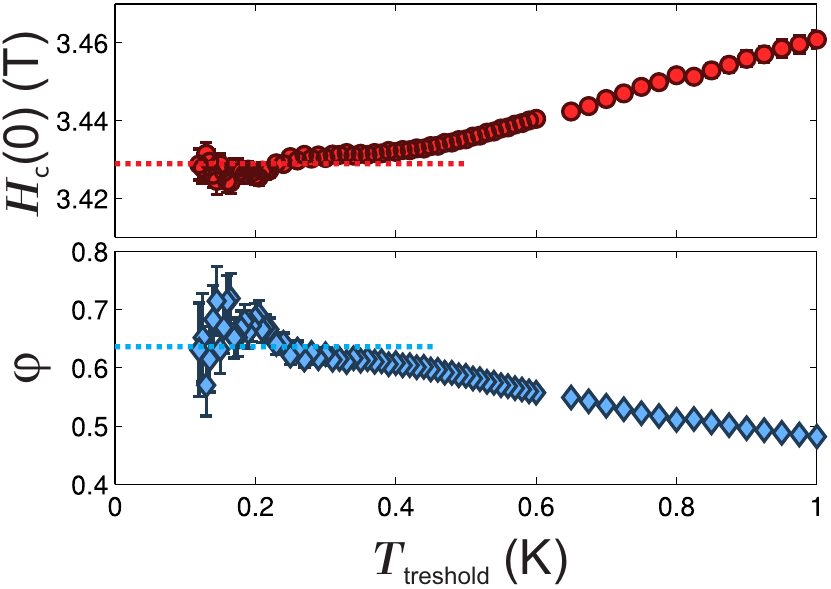}
        \caption{Windowing analysis of \sulf\ phase boundary obtained from dielectric measurements.}\label{FIG:window}
\end{center}
\end{figure}

To investigate the critical properties of the phase boundary, we
have performed a windowing analysis, similar to how it has been in
Ref.~\cite{Huvonen_PRB_2013_PHCXphasediagram}. The dependencies of
estimated critical field and crossover exponent $\varphi$ on the
cut-off temperature is shown in Fig.~\ref{FIG:window}. Around
$T\sim300$~mK the dependence of $H_{c}$ and $\varphi$ on the
threshold temperature vanishes, indicating the true power-law
behavior. For extremely narrow fit windows the resulting values are
slightly scattered due to the decreasing amount of datapoints. For a
wider fitting range $\lesssim1$~K, we get $\varphi\simeq0.5$, thus
recovering the value previously found by specific heat
measurements~\cite{Fujisawa_ProgThPhys_2005_SulfHC}.

 The observed critical
filed value is different from the one quoted in the introduction,
due to the direction of applied field in the present study
($\mathbf{H}||\mathbf{a}$) being different from that in previous
neutron \protect\cite{GarleaZheludev_PRB_2009_SulfOrdering} and
calorimetry \protect\cite{WulfMuhlbauer_PRB_2011_SulDisordered}
experiments ($\mathbf{H}||\mathbf{b}$). Preceding dielectric
study~\cite{Schrettle_PRB_2013_SulMF} also used
$\mathbf{H}||\mathbf{a}$ field orientation,and their phase boundary
is apparently shifted from the values for $\mathbf{H}||\mathbf{b}$
they quote in Fig.~4.

\section{Magnetocapacitive effect}

\subsection{Symmetry properties}

The spiral order in \sulf\ can be parameterized as
$\aver{\mathbf{S}_{\perp}(\mathbf{r})}=\mathbf{S}_{1}\cos(\mathbf{Qr})+\mathbf{S}_{2}\sin(\mathbf{Qr})$
with $\mathbf{S}_{1}\perp\mathbf{S}_{2}$. This spiral arrangement,
as it has been shown by Mostovoy~\cite{Mostovoy_PRL_2006_MFmacro}
and Katsura, Nagaosa and Balatsky~\cite{Katsura_PRL_2005_MFmicro}
induces an electric polarization
$\mathbf{P}\propto\boldsymbol{\left[}\left[\mathbf{S}_{1}\times\mathbf{S}_{2}\right]\times\mathbf{Q}\boldsymbol{\right]}$
due to lack of inversion symmetry between the spins and non-zero
spin-orbit coupling. This effect is often called ``inverse
Dzyaloshinskii--Moriya" mechanism, due to analogy with the
Dzyaloshinskii--Moriya antisymmetric
coupling~\cite{Dzyaloshinskii_JPChemS_1958_DM,Moriya_PR_1960_DM}. In
the latter case the asymmetry of the charge distribution between the
magnetic ions allows for a coupling term proportional to
$\left[\mathbf{S}_{1}\times\mathbf{S}_{2}\right]$, which may
stabilize a spiral phase. In the former case it is inverse: spiral
arrangement of spins enforces spatial asymmetry in charge
distribution.

\begin{figure}
\begin{center}
        \includegraphics[width=0.45\textwidth]{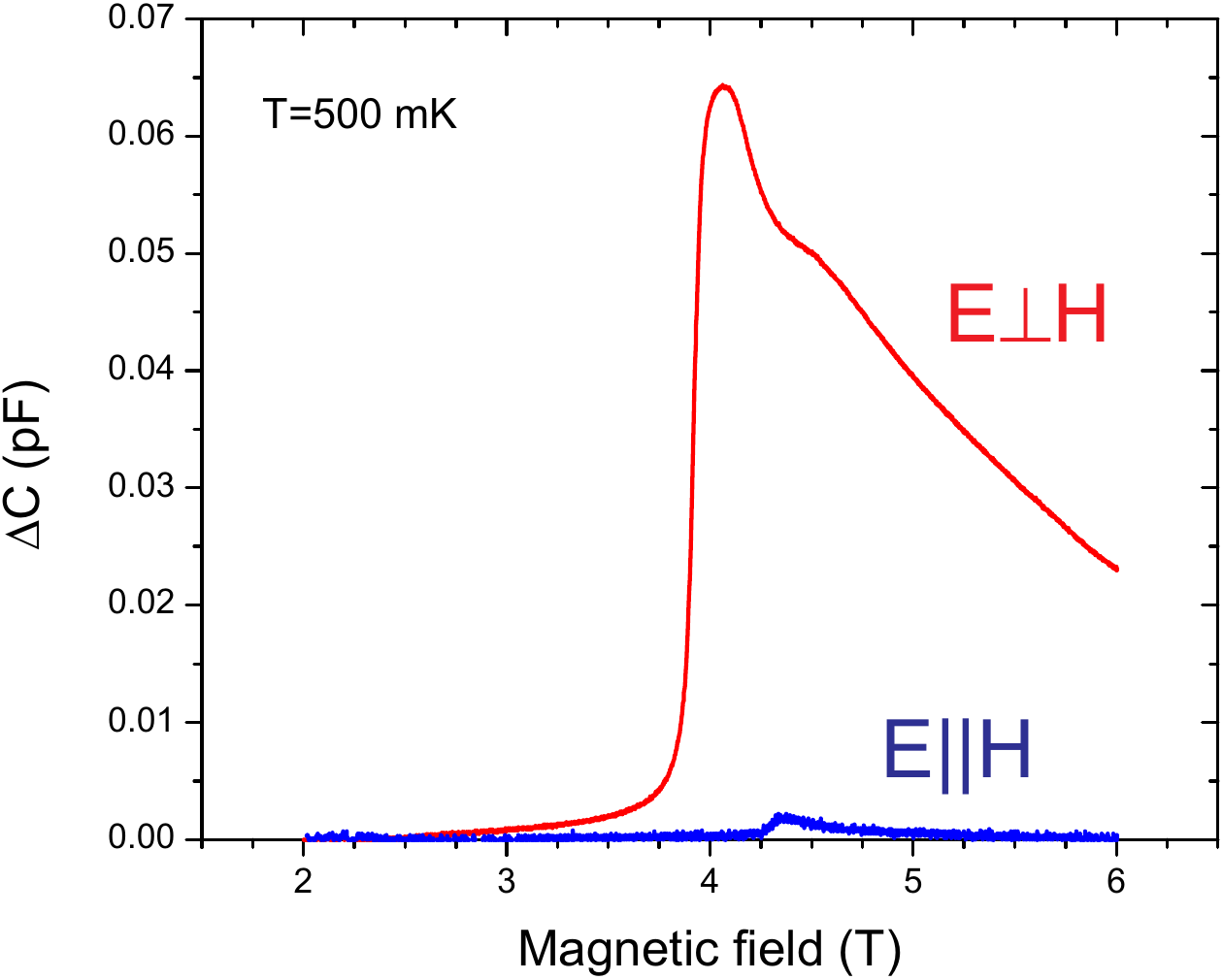}
        \caption{Magnetocapacitive effect in the same sample of \sulf, measured along ($\mathbf{E}\perp\mathbf{H}$) and perpendicular ($\mathbf{E}\parallel\mathbf{H}$) to the spiral plane.}\label{FIG:pp}
\end{center}
\end{figure}

The distinctive feature of this mechanism is the direction of
polarization, lying in the spiral plane and perpendicular to the
propagation vector $\mathbf{Q}$. Crystals of \sulf\ possess a
natural cleavage plane, which is almost parallel to $\mathbf{Q}$,
and the spiral plane is fixed by the external magnetic field, as
$\left[\mathbf{S}_{1}\times\mathbf{S}_{2}\right]\parallel
\mathbf{H}$. Hence, such a plane is ideally suited for the
observation of magnetocapacitive effect when
$\mathbf{H}\perp\mathbf{Q}$ and
$\mathbf{E}\perp\mathbf{Q},\mathbf{H}$ (i. e. $\mathbf{E}$
perpendicular to the ``good plane" and magnetic field also lies in
the ``good plane"). The bulk of the measurements was performed in
this crossed fields configuration. However, we also adopted an
alternative experimental setup: the same sample assembly was mounted
in parallel fields configuration, with
$\mathbf{H}\parallel\mathbf{E}$. This means, that the dielectric
response is probed perpendicular to the spiral plane. In such
configuration the magnetocapacitive effect is drastically suppressed
(see Fig.~\ref{FIG:pp}): while at $T=500$~mK a prominent peak is
seen in crossed fields configuration, only a tiny remnant peak can
be observed in the parallel fields configuration. This indicates
that the non-trivial dielectric response is restricted to the spiral
plane
--- exactly what should be expected for the ``inverse
Dzyaloshinskii--Moriya" mechanism.

\subsection{Domain-related effects}

\begin{figure}[b]
\begin{center}
        \includegraphics[width=0.5\textwidth]{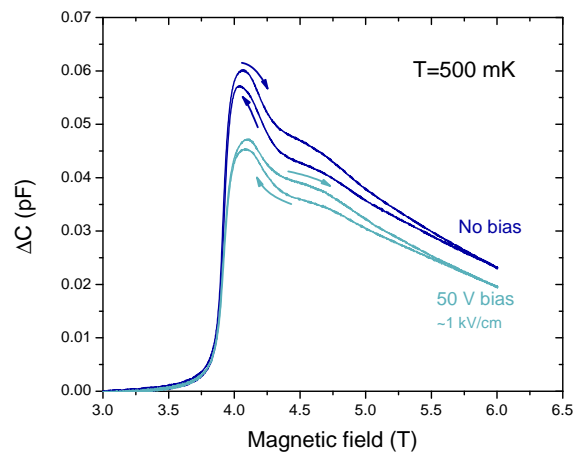}
        \caption{Examples of hysteretic behavior of magnetocapacitive effect in \sulf. Also note the suppression of the effect by bias field.}\label{FIG:bias}
\end{center}
\end{figure}

The magnetocapacitive effect, observed in \sulf, demonstrates a
history dependence in the magnetically ordered phase. The example
can be found in Fig.~\ref{FIG:bias}, where the field is swept from
$3$ to $6$~T and then back at rate of $10^{-3}$~T/sec. The
hysteresis onset is located above the inflection point of the steep
slope of $\Delta C(H)$. Going below the critical field ``resets''
the history, making the curves shown in Fig.~\ref{FIG:bias} fully
reproducible over multiple runs. Also, note the suppression of the
magnetocapacitive anomaly by a constant (bias) electric field
applied to the sample. All together this is a strong indication of
domain formation above $H_{c}$. In zero electric field one may
expect equal population of clockwise and counterclockwise spiral
domains, having the opposite direction of electric polarization
vector. Note that the spiral plane is fixed by external magnetic
field, and hence the polarization of the domains must behave in
Ising-like way. Having such domains can be a natural cause for the
memory effect in the ordered phase. Also, it explains the
suppression of the effect by bias electric field. Such field would
decrease the response of $P$ to the AC probing field, as it would
fix the preferred direction of domain polarization.

\bibliography{The_Library}

\begin{thebibliography}{33}%
\makeatletter
\providecommand \@ifxundefined [1]{%
 \@ifx{#1\undefined}
}%
\providecommand \@ifnum [1]{%
 \ifnum #1\expandafter \@firstoftwo
 \else \expandafter \@secondoftwo
 \fi
}%
\providecommand \@ifx [1]{%
 \ifx #1\expandafter \@firstoftwo
 \else \expandafter \@secondoftwo
 \fi
}%
\providecommand \natexlab [1]{#1}%
\providecommand \enquote  [1]{``#1''}%
\providecommand \bibnamefont  [1]{#1}%
\providecommand \bibfnamefont [1]{#1}%
\providecommand \citenamefont [1]{#1}%
\providecommand \href@noop [0]{\@secondoftwo}%
\providecommand \href [0]{\begingroup \@sanitize@url \@href}%
\providecommand \@href[1]{\@@startlink{#1}\@@href}%
\providecommand \@@href[1]{\endgroup#1\@@endlink}%
\providecommand \@sanitize@url [0]{\catcode `\\12\catcode `\$12\catcode
  `\&12\catcode `\#12\catcode `\^12\catcode `\_12\catcode `\%12\relax}%
\providecommand \@@startlink[1]{}%
\providecommand \@@endlink[0]{}%
\providecommand \url  [0]{\begingroup\@sanitize@url \@url }%
\providecommand \@url [1]{\endgroup\@href {#1}{\urlprefix }}%
\providecommand \urlprefix  [0]{URL }%
\providecommand \Eprint [0]{\href }%
\providecommand \doibase [0]{http://dx.doi.org/}%
\providecommand \selectlanguage [0]{\@gobble}%
\providecommand \bibinfo  [0]{\@secondoftwo}%
\providecommand \bibfield  [0]{\@secondoftwo}%
\providecommand \translation [1]{[#1]}%
\providecommand \BibitemOpen [0]{}%
\providecommand \bibitemStop [0]{}%
\providecommand \bibitemNoStop [0]{.\EOS\space}%
\providecommand \EOS [0]{\spacefactor3000\relax}%
\providecommand \BibitemShut  [1]{\csname bibitem#1\endcsname}%
\let\auto@bib@innerbib\@empty
\bibitem [{\citenamefont {Giamarchi}\ \emph {et~al.}(2008)\citenamefont
  {Giamarchi}, \citenamefont {R{\"u}egg},\ and\ \citenamefont
  {Tchernyshyov}}]{Giamarchi_NatPhys_2008_BECreview}%
  \BibitemOpen
  \bibfield  {author} {\bibinfo {author} {\bibfnamefont {T.}~\bibnamefont
  {Giamarchi}}, \bibinfo {author} {\bibfnamefont {C.}~\bibnamefont
  {R{\"u}egg}}, \ and\ \bibinfo {author} {\bibfnamefont {O.}~\bibnamefont
  {Tchernyshyov}},\ }\bibfield  {title} {\enquote {\bibinfo {title}
  {{Bose-Einstein condensation in magnetic insulators}},}\ }\href {\doibase
  10.1038/nphys893} {\bibfield  {journal} {\bibinfo  {journal} {Nat. Physics}\
  }\textbf {\bibinfo {volume} {4}},\ \bibinfo {pages} {198} (\bibinfo {year}
  {2008})}\BibitemShut {NoStop}%
\bibitem [{\citenamefont {Zapf}\ \emph {et~al.}(2014)\citenamefont {Zapf},
  \citenamefont {Jaime},\ and\ \citenamefont
  {Batista}}]{Zapf_RMP_2014_BECreview}%
  \BibitemOpen
  \bibfield  {author} {\bibinfo {author} {\bibfnamefont {V.}~\bibnamefont
  {Zapf}}, \bibinfo {author} {\bibfnamefont {M.}~\bibnamefont {Jaime}}, \ and\
  \bibinfo {author} {\bibfnamefont {C.~D.}\ \bibnamefont {Batista}},\
  }\bibfield  {title} {\enquote {\bibinfo {title} {{Bose-Einstein condensation
  in quantum magnets}},}\ }\href {\doibase 10.1103/RevModPhys.86.563}
  {\bibfield  {journal} {\bibinfo  {journal} {Rev. Mod. Phys.}\ }\textbf
  {\bibinfo {volume} {86}},\ \bibinfo {pages} {563} (\bibinfo {year}
  {2014})}\BibitemShut {NoStop}%
\bibitem [{\citenamefont {Cheong}\ and\ \citenamefont
  {Mostovoy}(2007)}]{CheongMosotovoy_NatMat_2007_NatureReview}%
  \BibitemOpen
  \bibfield  {author} {\bibinfo {author} {\bibfnamefont {S.-W.}\ \bibnamefont
  {Cheong}}\ and\ \bibinfo {author} {\bibfnamefont {M.}~\bibnamefont
  {Mostovoy}},\ }\bibfield  {title} {\enquote {\bibinfo {title}
  {{Multiferroics: a magnetic twist for ferroelectricity}},}\ }\href
  {http://www.nature.com/nmat/journal/v6/n1/abs/nmat1804.html} {\bibfield
  {journal} {\bibinfo  {journal} {Nat. Mater.}\ }\textbf {\bibinfo {volume}
  {6}},\ \bibinfo {pages} {13} (\bibinfo {year} {2007})}\BibitemShut {NoStop}%
\bibitem [{\citenamefont {Katsura}\ \emph {et~al.}(2005)\citenamefont
  {Katsura}, \citenamefont {Nagaosa},\ and\ \citenamefont
  {Balatsky}}]{Katsura_PRL_2005_MFmicro}%
  \BibitemOpen
  \bibfield  {author} {\bibinfo {author} {\bibfnamefont {H.}~\bibnamefont
  {Katsura}}, \bibinfo {author} {\bibfnamefont {N.}~\bibnamefont {Nagaosa}}, \
  and\ \bibinfo {author} {\bibfnamefont {A.~V.}\ \bibnamefont {Balatsky}},\
  }\bibfield  {title} {\enquote {\bibinfo {title} {Spin current and
  magnetoelectric effect in noncollinear magnets},}\ }\href {\doibase
  10.1103/PhysRevLett.95.057205} {\bibfield  {journal} {\bibinfo  {journal}
  {Phys. Rev. Lett.}\ }\textbf {\bibinfo {volume} {95}},\ \bibinfo {pages}
  {057205} (\bibinfo {year} {2005})}\BibitemShut {NoStop}%
\bibitem [{\citenamefont {Mostovoy}(2006)}]{Mostovoy_PRL_2006_MFmacro}%
  \BibitemOpen
  \bibfield  {author} {\bibinfo {author} {\bibfnamefont {M.}~\bibnamefont
  {Mostovoy}},\ }\bibfield  {title} {\enquote {\bibinfo {title}
  {Ferroelectricity in spiral magnets},}\ }\href {\doibase
  10.1103/PhysRevLett.96.067601} {\bibfield  {journal} {\bibinfo  {journal}
  {Phys. Rev. Lett.}\ }\textbf {\bibinfo {volume} {96}},\ \bibinfo {pages}
  {067601} (\bibinfo {year} {2006})}\BibitemShut {NoStop}%
\bibitem [{\citenamefont {Picozzi}\ \emph {et~al.}(2007)\citenamefont
  {Picozzi}, \citenamefont {Yamauchi}, \citenamefont {Sanyal}, \citenamefont
  {Sergienko},\ and\ \citenamefont
  {Dagotto}}]{SergienkoDagotto_PRL_2007_DualMF}%
  \BibitemOpen
  \bibfield  {author} {\bibinfo {author} {\bibfnamefont {S.}~\bibnamefont
  {Picozzi}}, \bibinfo {author} {\bibfnamefont {K.}~\bibnamefont {Yamauchi}},
  \bibinfo {author} {\bibfnamefont {B.}~\bibnamefont {Sanyal}}, \bibinfo
  {author} {\bibfnamefont {I.~A.}\ \bibnamefont {Sergienko}}, \ and\ \bibinfo
  {author} {\bibfnamefont {E.}~\bibnamefont {Dagotto}},\ }\bibfield  {title}
  {\enquote {\bibinfo {title} {Dual nature of improper ferroelectricity in a
  magnetoelectric multiferroic},}\ }\href {\doibase
  10.1103/PhysRevLett.99.227201} {\bibfield  {journal} {\bibinfo  {journal}
  {Phys. Rev. Lett.}\ }\textbf {\bibinfo {volume} {99}},\ \bibinfo {pages}
  {227201} (\bibinfo {year} {2007})}\BibitemShut {NoStop}%
\bibitem [{\citenamefont {Kenzelmann}\ \emph {et~al.}(2007)\citenamefont
  {Kenzelmann}, \citenamefont {Lawes}, \citenamefont {Harris}, \citenamefont
  {Gasparovic}, \citenamefont {Broholm}, \citenamefont {Ramirez}, \citenamefont
  {Jorge}, \citenamefont {Jaime}, \citenamefont {Park}, \citenamefont {Huang},
  \citenamefont {Shapiro},\ and\ \citenamefont
  {Demianets}}]{Kenzelmann_PRL_2007_TriangularFerro}%
  \BibitemOpen
  \bibfield  {author} {\bibinfo {author} {\bibfnamefont {M.}~\bibnamefont
  {Kenzelmann}}, \bibinfo {author} {\bibfnamefont {G.}~\bibnamefont {Lawes}},
  \bibinfo {author} {\bibfnamefont {A.~B.}\ \bibnamefont {Harris}}, \bibinfo
  {author} {\bibfnamefont {G.}~\bibnamefont {Gasparovic}}, \bibinfo {author}
  {\bibfnamefont {C.}~\bibnamefont {Broholm}}, \bibinfo {author} {\bibfnamefont
  {A.~P.}\ \bibnamefont {Ramirez}}, \bibinfo {author} {\bibfnamefont {G.~A.}\
  \bibnamefont {Jorge}}, \bibinfo {author} {\bibfnamefont {M.}~\bibnamefont
  {Jaime}}, \bibinfo {author} {\bibfnamefont {S.}~\bibnamefont {Park}},
  \bibinfo {author} {\bibfnamefont {Q.}~\bibnamefont {Huang}}, \bibinfo
  {author} {\bibfnamefont {A.~Ya.}\ \bibnamefont {Shapiro}}, \ and\ \bibinfo
  {author} {\bibfnamefont {L.~A.}\ \bibnamefont {Demianets}},\ }\bibfield
  {title} {\enquote {\bibinfo {title} {Direct transition from a disordered to a
  multiferroic phase on a triangular lattice},}\ }\href {\doibase
  10.1103/PhysRevLett.98.267205} {\bibfield  {journal} {\bibinfo  {journal}
  {Phys. Rev. Lett.}\ }\textbf {\bibinfo {volume} {98}},\ \bibinfo {pages}
  {267205} (\bibinfo {year} {2007})}\BibitemShut {NoStop}%
\bibitem [{\citenamefont {Abe}\ \emph {et~al.}(2009)\citenamefont {Abe},
  \citenamefont {Taniguchi}, \citenamefont {Ohtani}, \citenamefont {Umetsu},\
  and\ \citenamefont {Arima}}]{Abe_PRB_2009_TbMnO3Flop}%
  \BibitemOpen
  \bibfield  {author} {\bibinfo {author} {\bibfnamefont {N.}~\bibnamefont
  {Abe}}, \bibinfo {author} {\bibfnamefont {K.}~\bibnamefont {Taniguchi}},
  \bibinfo {author} {\bibfnamefont {S.}~\bibnamefont {Ohtani}}, \bibinfo
  {author} {\bibfnamefont {H.}~\bibnamefont {Umetsu}}, \ and\ \bibinfo {author}
  {\bibfnamefont {T.}~\bibnamefont {Arima}},\ }\bibfield  {title} {\enquote
  {\bibinfo {title} {{Control of the polarization flop direction by a tilted
  magnetic field in multiferroic ${\text{TbMnO}}_{3}$}},}\ }\href {\doibase
  10.1103/PhysRevB.80.020402} {\bibfield  {journal} {\bibinfo  {journal} {Phys.
  Rev. B}\ }\textbf {\bibinfo {volume} {80}},\ \bibinfo {pages} {020402}
  (\bibinfo {year} {2009})}\BibitemShut {NoStop}%
\bibitem [{\citenamefont {Kim}\ \emph {et~al.}(2014)\citenamefont {Kim},
  \citenamefont {Khim}, \citenamefont {Chun}, \citenamefont {Jo}, \citenamefont
  {Balicas}, \citenamefont {Yi}, \citenamefont {Cheong}, \citenamefont
  {Harrison}, \citenamefont {Batista}, \citenamefont {Han},\ and\ \citenamefont
  {Kim}}]{Kim_NatComm_2014_SaturationSusceptibility}%
  \BibitemOpen
  \bibfield  {author} {\bibinfo {author} {\bibfnamefont {J.~W.}\ \bibnamefont
  {Kim}}, \bibinfo {author} {\bibfnamefont {S.}~\bibnamefont {Khim}}, \bibinfo
  {author} {\bibfnamefont {S.~H.}\ \bibnamefont {Chun}}, \bibinfo {author}
  {\bibfnamefont {Y.}~\bibnamefont {Jo}}, \bibinfo {author} {\bibfnamefont
  {L.}~\bibnamefont {Balicas}}, \bibinfo {author} {\bibfnamefont {H.~T.}\
  \bibnamefont {Yi}}, \bibinfo {author} {\bibfnamefont {S.-W.}\ \bibnamefont
  {Cheong}}, \bibinfo {author} {\bibfnamefont {N.}~\bibnamefont {Harrison}},
  \bibinfo {author} {\bibfnamefont {C.~D.}\ \bibnamefont {Batista}}, \bibinfo
  {author} {\bibfnamefont {J.~H.}\ \bibnamefont {Han}}, \ and\ \bibinfo
  {author} {\bibfnamefont {K.~H.}\ \bibnamefont {Kim}},\ }\bibfield  {title}
  {\enquote {\bibinfo {title} {Manifestation of magnetic quantum fluctuations
  in the dielectric properties of a multiferroic},}\ }\href {\doibase
  10.1038/ncomms5419} {\bibfield  {journal} {\bibinfo  {journal} {Nat.
  Commun.}\ }\textbf {\bibinfo {volume} {5}},\ \bibinfo {pages} {4419}
  (\bibinfo {year} {2014})}\BibitemShut {NoStop}%
\bibitem [{\citenamefont {Fujisawa}\ \emph {et~al.}(2005)\citenamefont
  {Fujisawa}, \citenamefont {Tanaka},\ and\ \citenamefont
  {Sakakibara}}]{Fujisawa_ProgThPhys_2005_SulfHC}%
  \BibitemOpen
  \bibfield  {author} {\bibinfo {author} {\bibfnamefont {M.}~\bibnamefont
  {Fujisawa}}, \bibinfo {author} {\bibfnamefont {H.}~\bibnamefont {Tanaka}}, \
  and\ \bibinfo {author} {\bibfnamefont {T.}~\bibnamefont {Sakakibara}},\
  }\bibfield  {title} {\enquote {\bibinfo {title} {Magnetic field-induced phase
  transition in quantum spin system
  {Cu$_2$Cl$_4${\textperiodcentered}H$_8$C$_4$SO$_2$}},}\ }\href {\doibase
  10.1143/PTPS.159.212} {\bibfield  {journal} {\bibinfo  {journal} {Prog.
  Theor. Phys. Supp.}\ }\textbf {\bibinfo {volume} {159}},\ \bibinfo {pages}
  {212} (\bibinfo {year} {2005})}\BibitemShut {NoStop}%
\bibitem [{\citenamefont {Garlea}\ \emph {et~al.}(2008)\citenamefont {Garlea},
  \citenamefont {Zheludev}, \citenamefont {Regnault}, \citenamefont {Chung},
  \citenamefont {Qiu}, \citenamefont {Boehm}, \citenamefont {Habicht},\ and\
  \citenamefont {Meissner}}]{GarleaZheludev_PRL_2008_SulfZF}%
  \BibitemOpen
  \bibfield  {author} {\bibinfo {author} {\bibfnamefont {V.~O.}\ \bibnamefont
  {Garlea}}, \bibinfo {author} {\bibfnamefont {A.}~\bibnamefont {Zheludev}},
  \bibinfo {author} {\bibfnamefont {L.-P.}\ \bibnamefont {Regnault}}, \bibinfo
  {author} {\bibfnamefont {J.-H.}\ \bibnamefont {Chung}}, \bibinfo {author}
  {\bibfnamefont {Y.}~\bibnamefont {Qiu}}, \bibinfo {author} {\bibfnamefont
  {M.}~\bibnamefont {Boehm}}, \bibinfo {author} {\bibfnamefont
  {K.}~\bibnamefont {Habicht}}, \ and\ \bibinfo {author} {\bibfnamefont
  {M.}~\bibnamefont {Meissner}},\ }\bibfield  {title} {\enquote {\bibinfo
  {title} {{Excitations in a Four-Leg Antiferromagnetic Heisenberg Spin
  Tube}},}\ }\href {\doibase 10.1103/PhysRevLett.100.037206} {\bibfield
  {journal} {\bibinfo  {journal} {Phys. Rev. Lett.}\ }\textbf {\bibinfo
  {volume} {100}},\ \bibinfo {pages} {037206} (\bibinfo {year}
  {2008})}\BibitemShut {NoStop}%
\bibitem [{\citenamefont {Garlea}\ \emph {et~al.}(2009)\citenamefont {Garlea},
  \citenamefont {Zheludev}, \citenamefont {Habicht}, \citenamefont {Meissner},
  \citenamefont {Grenier}, \citenamefont {Regnault},\ and\ \citenamefont
  {Ressouche}}]{GarleaZheludev_PRB_2009_SulfOrdering}%
  \BibitemOpen
  \bibfield  {author} {\bibinfo {author} {\bibfnamefont {V.~O.}\ \bibnamefont
  {Garlea}}, \bibinfo {author} {\bibfnamefont {A.}~\bibnamefont {Zheludev}},
  \bibinfo {author} {\bibfnamefont {K.}~\bibnamefont {Habicht}}, \bibinfo
  {author} {\bibfnamefont {M.}~\bibnamefont {Meissner}}, \bibinfo {author}
  {\bibfnamefont {B.}~\bibnamefont {Grenier}}, \bibinfo {author} {\bibfnamefont
  {L.-P.}\ \bibnamefont {Regnault}}, \ and\ \bibinfo {author} {\bibfnamefont
  {E.}~\bibnamefont {Ressouche}},\ }\bibfield  {title} {\enquote {\bibinfo
  {title} {Dimensional crossover in a spin-liquid-to-helimagnet quantum phase
  transition},}\ }\href {\doibase 10.1103/PhysRevB.79.060404} {\bibfield
  {journal} {\bibinfo  {journal} {Phys. Rev. B}\ }\textbf {\bibinfo {volume}
  {79}},\ \bibinfo {pages} {060404} (\bibinfo {year} {2009})}\BibitemShut
  {NoStop}%
\bibitem [{\citenamefont {Zheludev}\ \emph {et~al.}(2009)\citenamefont
  {Zheludev}, \citenamefont {Garlea}, \citenamefont {Tsvelik}, \citenamefont
  {Regnault}, \citenamefont {Habicht}, \citenamefont {Kiefer},\ and\
  \citenamefont {Roessli}}]{ZheludevGarlea_PRB_2009_SulfExcitations}%
  \BibitemOpen
  \bibfield  {author} {\bibinfo {author} {\bibfnamefont {A.}~\bibnamefont
  {Zheludev}}, \bibinfo {author} {\bibfnamefont {V.~O.}\ \bibnamefont
  {Garlea}}, \bibinfo {author} {\bibfnamefont {A.}~\bibnamefont {Tsvelik}},
  \bibinfo {author} {\bibfnamefont {L.-P.}\ \bibnamefont {Regnault}}, \bibinfo
  {author} {\bibfnamefont {K.}~\bibnamefont {Habicht}}, \bibinfo {author}
  {\bibfnamefont {K.}~\bibnamefont {Kiefer}}, \ and\ \bibinfo {author}
  {\bibfnamefont {B.}~\bibnamefont {Roessli}},\ }\bibfield  {title} {\enquote
  {\bibinfo {title} {Excitations from a chiral magnetized state of a frustrated
  quantum spin liquid},}\ }\href {\doibase 10.1103/PhysRevB.80.214413}
  {\bibfield  {journal} {\bibinfo  {journal} {Phys. Rev. B}\ }\textbf {\bibinfo
  {volume} {80}},\ \bibinfo {pages} {214413} (\bibinfo {year}
  {2009})}\BibitemShut {NoStop}%
\bibitem [{\citenamefont {Schrettle}\ \emph {et~al.}(2013)\citenamefont
  {Schrettle}, \citenamefont {Krohns}, \citenamefont {Lunkenheimer},
  \citenamefont {Loidl}, \citenamefont {Wulf}, \citenamefont {Yankova},\ and\
  \citenamefont {Zheludev}}]{Schrettle_PRB_2013_SulMF}%
  \BibitemOpen
  \bibfield  {author} {\bibinfo {author} {\bibfnamefont {F.}~\bibnamefont
  {Schrettle}}, \bibinfo {author} {\bibfnamefont {S.}~\bibnamefont {Krohns}},
  \bibinfo {author} {\bibfnamefont {P.}~\bibnamefont {Lunkenheimer}}, \bibinfo
  {author} {\bibfnamefont {A.}~\bibnamefont {Loidl}}, \bibinfo {author}
  {\bibfnamefont {E.}~\bibnamefont {Wulf}}, \bibinfo {author} {\bibfnamefont
  {T.}~\bibnamefont {Yankova}}, \ and\ \bibinfo {author} {\bibfnamefont
  {A.}~\bibnamefont {Zheludev}},\ }\bibfield  {title} {\enquote {\bibinfo
  {title} {Magnetic-field induced multiferroicity in a quantum critical
  frustrated spin liquid},}\ }\href {\doibase 10.1103/PhysRevB.87.121105}
  {\bibfield  {journal} {\bibinfo  {journal} {Phys. Rev. B}\ }\textbf {\bibinfo
  {volume} {87}},\ \bibinfo {pages} {121105} (\bibinfo {year}
  {2013})}\BibitemShut {NoStop}%
\bibitem [{Sup()}]{SuppMat}%
  \BibitemOpen
  \href@noop {} {\bibinfo  {journal} {See Supplemental Material for more
  details}\ }\BibitemShut {NoStop}%
\bibitem [{\citenamefont
  {Dvo{\v{r}}{\'a}k}(1974)}]{Dvorak_Ferr_1974_improperReview}%
  \BibitemOpen
\bibfield  {journal} {  }\bibfield  {author} {\bibinfo {author} {\bibfnamefont
  {V.}~\bibnamefont {Dvo{\v{r}}{\'a}k}},\ }\bibfield  {title} {\enquote
  {\bibinfo {title} {Improper ferroelectrics},}\ }\href
  {http://www.tandfonline.com/doi/abs/10.1080/00150197408237942} {\bibfield
  {journal} {\bibinfo  {journal} {Ferroelectrics}\ }\textbf {\bibinfo {volume}
  {7}},\ \bibinfo {pages} {1} (\bibinfo {year} {1974})}\BibitemShut {NoStop}%
\bibitem [{\citenamefont {Wulf}\ \emph {et~al.}(2011)\citenamefont {Wulf},
  \citenamefont {M\"uhlbauer}, \citenamefont {Yankova},\ and\ \citenamefont
  {Zheludev}}]{WulfMuhlbauer_PRB_2011_SulDisordered}%
  \BibitemOpen
  \bibfield  {author} {\bibinfo {author} {\bibfnamefont {E.}~\bibnamefont
  {Wulf}}, \bibinfo {author} {\bibfnamefont {S.}~\bibnamefont {M\"uhlbauer}},
  \bibinfo {author} {\bibfnamefont {T.}~\bibnamefont {Yankova}}, \ and\
  \bibinfo {author} {\bibfnamefont {A.}~\bibnamefont {Zheludev}},\ }\bibfield
  {title} {\enquote {\bibinfo {title} {Disorder instability of the magnon
  condensate in a frustrated spin ladder},}\ }\href {\doibase
  10.1103/PhysRevB.84.174414} {\bibfield  {journal} {\bibinfo  {journal} {Phys.
  Rev. B}\ }\textbf {\bibinfo {volume} {84}},\ \bibinfo {pages} {174414}
  (\bibinfo {year} {2011})}\BibitemShut {NoStop}%
\bibitem [{\citenamefont {Dell'Amore}\ \emph {et~al.}(2009)\citenamefont
  {Dell'Amore}, \citenamefont {Schilling},\ and\ \citenamefont
  {Kr\"amer}}]{DellAmore_PRB_2009_BECbreakdown}%
  \BibitemOpen
  \bibfield  {author} {\bibinfo {author} {\bibfnamefont {R.}~\bibnamefont
  {Dell'Amore}}, \bibinfo {author} {\bibfnamefont {A.}~\bibnamefont
  {Schilling}}, \ and\ \bibinfo {author} {\bibfnamefont {K.}~\bibnamefont
  {Kr\"amer}},\ }\bibfield  {title} {\enquote {\bibinfo {title} {{$U(1)$
  symmetry breaking and violated axial symmetry in ${\text{TlCuCl}}_{3}$ and
  other insulating spin systems}},}\ }\href {\doibase
  10.1103/PhysRevB.79.014438} {\bibfield  {journal} {\bibinfo  {journal} {Phys.
  Rev. B}\ }\textbf {\bibinfo {volume} {79}},\ \bibinfo {pages} {014438}
  (\bibinfo {year} {2009})}\BibitemShut {NoStop}%
\bibitem [{\citenamefont {Glazkov}\ \emph {et~al.}(2004)\citenamefont
  {Glazkov}, \citenamefont {Smirnov}, \citenamefont {Tanaka},\ and\
  \citenamefont {Oosawa}}]{Glazkov_PRB_2004_TlCuCl3gap}%
  \BibitemOpen
  \bibfield  {author} {\bibinfo {author} {\bibfnamefont {V.~N.}\ \bibnamefont
  {Glazkov}}, \bibinfo {author} {\bibfnamefont {A.~I.}\ \bibnamefont
  {Smirnov}}, \bibinfo {author} {\bibfnamefont {H.}~\bibnamefont {Tanaka}}, \
  and\ \bibinfo {author} {\bibfnamefont {A.}~\bibnamefont {Oosawa}},\
  }\bibfield  {title} {\enquote {\bibinfo {title} {{Spin-resonance modes of the
  spin-gap magnet ${\mathrm{TlCuCl}}_{3}$}},}\ }\href {\doibase
  10.1103/PhysRevB.69.184410} {\bibfield  {journal} {\bibinfo  {journal} {Phys.
  Rev. B}\ }\textbf {\bibinfo {volume} {69}},\ \bibinfo {pages} {184410}
  (\bibinfo {year} {2004})}\BibitemShut {NoStop}%
\bibitem [{\citenamefont {Sirker}\ \emph {et~al.}(2004)\citenamefont {Sirker},
  \citenamefont {Wei{\ss}e},\ and\ \citenamefont
  {Sushkov}}]{Sirker_EurPLett_2004_SOCBEC}%
  \BibitemOpen
  \bibfield  {author} {\bibinfo {author} {\bibfnamefont {J.}~\bibnamefont
  {Sirker}}, \bibinfo {author} {\bibfnamefont {A.}~\bibnamefont {Wei{\ss}e}}, \
  and\ \bibinfo {author} {\bibfnamefont {O.~P.}\ \bibnamefont {Sushkov}},\
  }\bibfield  {title} {\enquote {\bibinfo {title} {{Consequences of spin-orbit
  coupling for the Bose-Einstein condensation of magnons}},}\ }\href {\doibase
  10.1209/epl/i2004-10179-4} {\bibfield  {journal} {\bibinfo  {journal}
  {Europhys. Lett.}\ }\textbf {\bibinfo {volume} {68}},\ \bibinfo {pages} {275}
  (\bibinfo {year} {2004})}\BibitemShut {NoStop}%
\bibitem [{\citenamefont {Zapf}\ \emph {et~al.}(2006)\citenamefont {Zapf},
  \citenamefont {Zocco}, \citenamefont {Hansen}, \citenamefont {Jaime},
  \citenamefont {Harrison}, \citenamefont {Batista}, \citenamefont
  {Kenzelmann}, \citenamefont {Niedermayer}, \citenamefont {Lacerda},\ and\
  \citenamefont {Paduan-Filho}}]{Zapf_PRL_2006_BECinDTN}%
  \BibitemOpen
  \bibfield  {author} {\bibinfo {author} {\bibfnamefont {V.~S.}\ \bibnamefont
  {Zapf}}, \bibinfo {author} {\bibfnamefont {D.}~\bibnamefont {Zocco}},
  \bibinfo {author} {\bibfnamefont {B.~R.}\ \bibnamefont {Hansen}}, \bibinfo
  {author} {\bibfnamefont {M.}~\bibnamefont {Jaime}}, \bibinfo {author}
  {\bibfnamefont {N.}~\bibnamefont {Harrison}}, \bibinfo {author}
  {\bibfnamefont {C.~D.}\ \bibnamefont {Batista}}, \bibinfo {author}
  {\bibfnamefont {M.}~\bibnamefont {Kenzelmann}}, \bibinfo {author}
  {\bibfnamefont {C.}~\bibnamefont {Niedermayer}}, \bibinfo {author}
  {\bibfnamefont {A.}~\bibnamefont {Lacerda}}, \ and\ \bibinfo {author}
  {\bibfnamefont {A.}~\bibnamefont {Paduan-Filho}},\ }\bibfield  {title}
  {\enquote {\bibinfo {title} {{Bose-Einstein Condensation of $S=1$ Nickel Spin
  Degrees of Freedom in
  ${\mathrm{NiCl}}_{2}\mathrm{\text{-}}4\mathrm{SC}({\mathrm{NH}}_{2}{)}_{2}$}},}\
  }\href {\doibase 10.1103/PhysRevLett.96.077204} {\bibfield  {journal}
  {\bibinfo  {journal} {Phys. Rev. Lett.}\ }\textbf {\bibinfo {volume} {96}},\
  \bibinfo {pages} {077204} (\bibinfo {year} {2006})}\BibitemShut {NoStop}%
\bibitem [{\citenamefont {Yin}\ \emph {et~al.}(2008)\citenamefont {Yin},
  \citenamefont {Xia}, \citenamefont {Zapf}, \citenamefont {Sullivan},\ and\
  \citenamefont {Paduan-Filho}}]{YinXia_PRL_2008_DTNcritical}%
  \BibitemOpen
  \bibfield  {author} {\bibinfo {author} {\bibfnamefont {L.}~\bibnamefont
  {Yin}}, \bibinfo {author} {\bibfnamefont {J.~S.}\ \bibnamefont {Xia}},
  \bibinfo {author} {\bibfnamefont {V.~S.}\ \bibnamefont {Zapf}}, \bibinfo
  {author} {\bibfnamefont {N.~S.}\ \bibnamefont {Sullivan}}, \ and\ \bibinfo
  {author} {\bibfnamefont {A.}~\bibnamefont {Paduan-Filho}},\ }\bibfield
  {title} {\enquote {\bibinfo {title} {{Direct Measurement of the Bose-Einstein
  Condensation Universality Class in
  ${\mathrm{NiCl}}_{2}\mathrm{\text{-}}4\mathrm{SC}({\mathrm{NH}}_{2}{)}_{2}$
  at Ultralow Temperatures}},}\ }\href {\doibase
  10.1103/PhysRevLett.101.187205} {\bibfield  {journal} {\bibinfo  {journal}
  {Phys. Rev. Lett.}\ }\textbf {\bibinfo {volume} {101}},\ \bibinfo {pages}
  {187205} (\bibinfo {year} {2008})}\BibitemShut {NoStop}%
\bibitem [{\citenamefont {Wulf}\ \emph {et~al.}(2015)\citenamefont {Wulf},
  \citenamefont {H\"{u}vonen}, \citenamefont {Sch\"{o}nemann}, \citenamefont
  {K\"{u}hne}, \citenamefont {Herrmannsd\"{o}rfer}, \citenamefont {Glavatskyy},
  \citenamefont {Gerischer}, \citenamefont {Kiefer}, \citenamefont
  {Gvasaliya},\ and\ \citenamefont
  {Zheludev}}]{WulfHuvonen_PRB_2015_DTNIntrinsicBroadening}%
  \BibitemOpen
  \bibfield  {author} {\bibinfo {author} {\bibfnamefont {E.}~\bibnamefont
  {Wulf}}, \bibinfo {author} {\bibfnamefont {D.}~\bibnamefont {H\"{u}vonen}},
  \bibinfo {author} {\bibfnamefont {R.}~\bibnamefont {Sch\"{o}nemann}},
  \bibinfo {author} {\bibfnamefont {H.}~\bibnamefont {K\"{u}hne}}, \bibinfo
  {author} {\bibfnamefont {T.}~\bibnamefont {Herrmannsd\"{o}rfer}}, \bibinfo
  {author} {\bibfnamefont {I.}~\bibnamefont {Glavatskyy}}, \bibinfo {author}
  {\bibfnamefont {S.}~\bibnamefont {Gerischer}}, \bibinfo {author}
  {\bibfnamefont {K.}~\bibnamefont {Kiefer}}, \bibinfo {author} {\bibfnamefont
  {S.}~\bibnamefont {Gvasaliya}}, \ and\ \bibinfo {author} {\bibfnamefont
  {A.}~\bibnamefont {Zheludev}},\ }\bibfield  {title} {\enquote {\bibinfo
  {title} {{Critical exponents and intrinsic broadening of the field-induced
  transition in
  ${\mathrm{NiCl}}_{2}\mathrm{\text{-}}4\mathrm{SC}({\mathrm{NH}}_{2}{)}_{2}$}},}\
  }\href {\doibase 10.1103/PhysRevB.91.014406} {\bibfield  {journal} {\bibinfo
  {journal} {Phys. Rev. B}\ }\textbf {\bibinfo {volume} {91}},\ \bibinfo
  {pages} {014406} (\bibinfo {year} {2015})}\BibitemShut {NoStop}%
\bibitem [{\citenamefont {Bruce}\ and\ \citenamefont
  {Cowley}(1978)}]{BruceCowley_JPC_1978_Phason}%
  \BibitemOpen
  \bibfield  {author} {\bibinfo {author} {\bibfnamefont {A.~D.}\ \bibnamefont
  {Bruce}}\ and\ \bibinfo {author} {\bibfnamefont {R.~A.}\ \bibnamefont
  {Cowley}},\ }\bibfield  {title} {\enquote {\bibinfo {title} {{The theory of
  structurally incommensurate systems. III. The fluctuation spectrum of
  incommensurate phases}},}\ }\href {\doibase 10.1088/0022-3719/11/17/014}
  {\bibfield  {journal} {\bibinfo  {journal} {J. Phys. C: Solid State Phys.}\
  }\textbf {\bibinfo {volume} {11}},\ \bibinfo {pages} {3609} (\bibinfo {year}
  {1978})}\BibitemShut {NoStop}%
\bibitem [{\citenamefont {Miga}\ \emph {et~al.}(2007)\citenamefont {Miga},
  \citenamefont {Dec},\ and\ \citenamefont
  {Kleemann}}]{Miga_RevSciIns_2007_DielectricSpectrometer}%
  \BibitemOpen
  \bibfield  {author} {\bibinfo {author} {\bibfnamefont {S.}~\bibnamefont
  {Miga}}, \bibinfo {author} {\bibfnamefont {J.}~\bibnamefont {Dec}}, \ and\
  \bibinfo {author} {\bibfnamefont {W.}~\bibnamefont {Kleemann}},\ }\bibfield
  {title} {\enquote {\bibinfo {title} {Computer-controlled susceptometer for
  investigating the linear and nonlinear dielectric response},}\ }\href
  {http://scitation.aip.org/content/aip/journal/rsi/78/3/10.1063/1.2712792}
  {\bibfield  {journal} {\bibinfo  {journal} {Rev. Sci. Inst.}\ }\textbf
  {\bibinfo {volume} {78}},\ \bibinfo {pages} {033902} (\bibinfo {year}
  {2007})}\BibitemShut {NoStop}%
\bibitem [{\citenamefont {Dec}\ \emph {et~al.}(2011)\citenamefont {Dec},
  \citenamefont {Miga},\ and\ \citenamefont
  {Kleemann}}]{Dec_Ferr_2011_DielectricNonlinReview}%
  \BibitemOpen
  \bibfield  {author} {\bibinfo {author} {\bibfnamefont {J.}~\bibnamefont
  {Dec}}, \bibinfo {author} {\bibfnamefont {S.}~\bibnamefont {Miga}}, \ and\
  \bibinfo {author} {\bibfnamefont {W.}~\bibnamefont {Kleemann}},\ }\bibfield
  {title} {\enquote {\bibinfo {title} {Ferroelectric phase transitions viewed
  via nonlinear dielectric response},}\ }\href
  {http://www.tandfonline.com/doi/abs/10.1080/00150193.2011.578500} {\bibfield
  {journal} {\bibinfo  {journal} {Ferroelectrics}\ }\textbf {\bibinfo {volume}
  {417}},\ \bibinfo {pages} {82} (\bibinfo {year} {2011})}\BibitemShut
  {NoStop}%
\bibitem [{\citenamefont {Seki}\ \emph {et~al.}(2010)\citenamefont {Seki},
  \citenamefont {Kurumaji}, \citenamefont {Ishiwata}, \citenamefont {Matsui},
  \citenamefont {Murakawa}, \citenamefont {Tokunaga}, \citenamefont {Kaneko},
  \citenamefont {Hasegawa},\ and\ \citenamefont
  {Tokura}}]{Seki_PRB_2010_CuCl2Multiferr}%
  \BibitemOpen
  \bibfield  {author} {\bibinfo {author} {\bibfnamefont {S.}~\bibnamefont
  {Seki}}, \bibinfo {author} {\bibfnamefont {T.}~\bibnamefont {Kurumaji}},
  \bibinfo {author} {\bibfnamefont {S.}~\bibnamefont {Ishiwata}}, \bibinfo
  {author} {\bibfnamefont {H.}~\bibnamefont {Matsui}}, \bibinfo {author}
  {\bibfnamefont {H.}~\bibnamefont {Murakawa}}, \bibinfo {author}
  {\bibfnamefont {Y.}~\bibnamefont {Tokunaga}}, \bibinfo {author}
  {\bibfnamefont {Y.}~\bibnamefont {Kaneko}}, \bibinfo {author} {\bibfnamefont
  {T.}~\bibnamefont {Hasegawa}}, \ and\ \bibinfo {author} {\bibfnamefont
  {Y.}~\bibnamefont {Tokura}},\ }\bibfield  {title} {\enquote {\bibinfo {title}
  {{Cupric chloride ${\text{CuCl}}_{2}$ as an $S=\frac{1}{2}$ chain
  multiferroic}},}\ }\href {\doibase 10.1103/PhysRevB.82.064424} {\bibfield
  {journal} {\bibinfo  {journal} {Phys. Rev. B}\ }\textbf {\bibinfo {volume}
  {82}},\ \bibinfo {pages} {064424} (\bibinfo {year} {2010})}\BibitemShut
  {NoStop}%
\bibitem [{\citenamefont {Park}\ \emph {et~al.}(2007)\citenamefont {Park},
  \citenamefont {Choi}, \citenamefont {Zhang},\ and\ \citenamefont
  {Cheong}}]{Park_PRL_2007_LiCu2O2Multiferr}%
  \BibitemOpen
  \bibfield  {author} {\bibinfo {author} {\bibfnamefont {S.}~\bibnamefont
  {Park}}, \bibinfo {author} {\bibfnamefont {Y.~J.}\ \bibnamefont {Choi}},
  \bibinfo {author} {\bibfnamefont {C.~L.}\ \bibnamefont {Zhang}}, \ and\
  \bibinfo {author} {\bibfnamefont {S-W.}\ \bibnamefont {Cheong}},\ }\bibfield
  {title} {\enquote {\bibinfo {title} {Ferroelectricity in an $s=1/2$ chain
  cuprate},}\ }\href {\doibase 10.1103/PhysRevLett.98.057601} {\bibfield
  {journal} {\bibinfo  {journal} {Phys. Rev. Lett.}\ }\textbf {\bibinfo
  {volume} {98}},\ \bibinfo {pages} {057601} (\bibinfo {year}
  {2007})}\BibitemShut {NoStop}%
\bibitem [{\citenamefont {Mourigal}\ \emph {et~al.}(2011)\citenamefont
  {Mourigal}, \citenamefont {Enderle}, \citenamefont {Kremer}, \citenamefont
  {Law},\ and\ \citenamefont {F\aa{}k}}]{Mourigal_PRB_2011_LiCuVO4spincurrent}%
  \BibitemOpen
  \bibfield  {author} {\bibinfo {author} {\bibfnamefont {M.}~\bibnamefont
  {Mourigal}}, \bibinfo {author} {\bibfnamefont {M.}~\bibnamefont {Enderle}},
  \bibinfo {author} {\bibfnamefont {R.~K.}\ \bibnamefont {Kremer}}, \bibinfo
  {author} {\bibfnamefont {J.~M.}\ \bibnamefont {Law}}, \ and\ \bibinfo
  {author} {\bibfnamefont {B.}~\bibnamefont {F\aa{}k}},\ }\bibfield  {title}
  {\enquote {\bibinfo {title} {{Ferroelectricity from spin supercurrents in
  LiCuVO${}_{4}$}},}\ }\href {\doibase 10.1103/PhysRevB.83.100409} {\bibfield
  {journal} {\bibinfo  {journal} {Phys. Rev. B}\ }\textbf {\bibinfo {volume}
  {83}},\ \bibinfo {pages} {100409} (\bibinfo {year} {2011})}\BibitemShut
  {NoStop}%
\bibitem [{\citenamefont {Ruff}\ \emph {et~al.}(2014)\citenamefont {Ruff},
  \citenamefont {Krohns}, \citenamefont {Lunkenheimer}, \citenamefont
  {Prokofiev},\ and\ \citenamefont
  {Loidl}}]{Ruff_JPCM_2014_LiCuVO4DomainFlops}%
  \BibitemOpen
  \bibfield  {author} {\bibinfo {author} {\bibfnamefont {A.}~\bibnamefont
  {Ruff}}, \bibinfo {author} {\bibfnamefont {S.}~\bibnamefont {Krohns}},
  \bibinfo {author} {\bibfnamefont {P.}~\bibnamefont {Lunkenheimer}}, \bibinfo
  {author} {\bibfnamefont {A.}~\bibnamefont {Prokofiev}}, \ and\ \bibinfo
  {author} {\bibfnamefont {A.}~\bibnamefont {Loidl}},\ }\bibfield  {title}
  {\enquote {\bibinfo {title} {{Dielectric properties and electrical switching
  behaviour of the spin-driven multiferroic LiCuVO$_4$}},}\ }\href {\doibase
  10.1088/0953-8984/26/48/485901} {\bibfield  {journal} {\bibinfo  {journal}
  {J. Phys.: Condens. Matter}\ }\textbf {\bibinfo {volume} {26}},\ \bibinfo
  {pages} {485901} (\bibinfo {year} {2014})}\BibitemShut {NoStop}%
\bibitem [{\citenamefont {H\"{u}vonen}\ \emph {et~al.}(2013)\citenamefont
  {H\"{u}vonen}, \citenamefont {Ballon},\ and\ \citenamefont
  {Zheludev}}]{Huvonen_PRB_2013_PHCXphasediagram}%
  \BibitemOpen
  \bibfield  {author} {\bibinfo {author} {\bibfnamefont {D.}~\bibnamefont
  {H\"{u}vonen}}, \bibinfo {author} {\bibfnamefont {G.}~\bibnamefont {Ballon}},
  \ and\ \bibinfo {author} {\bibfnamefont {A.}~\bibnamefont {Zheludev}},\
  }\bibfield  {title} {\enquote {\bibinfo {title} {Field-concentration phase
  diagram of a quantum spin liquid with bond defects},}\ }\href {\doibase
  10.1103/PhysRevB.88.094402} {\bibfield  {journal} {\bibinfo  {journal} {Phys.
  Rev. B}\ }\textbf {\bibinfo {volume} {88}},\ \bibinfo {pages} {094402}
  (\bibinfo {year} {2013})}\BibitemShut {NoStop}%
\bibitem [{\citenamefont
  {Dzyaloshinsky}(1958)}]{Dzyaloshinskii_JPChemS_1958_DM}%
  \BibitemOpen
  \bibfield  {author} {\bibinfo {author} {\bibfnamefont {I.}~\bibnamefont
  {Dzyaloshinsky}},\ }\bibfield  {title} {\enquote {\bibinfo {title} {{A
  thermodynamic theory of 'weak' ferromagnetism of antiferromagnetics}},}\
  }\href {\doibase 10.1016/0022-3697(58)90076-3} {\bibfield  {journal}
  {\bibinfo  {journal} {J. Phys. Chem. Solids}\ }\textbf {\bibinfo {volume}
  {4}},\ \bibinfo {pages} {241} (\bibinfo {year} {1958})}\BibitemShut {NoStop}%
\bibitem [{\citenamefont {Moriya}(1960)}]{Moriya_PR_1960_DM}%
  \BibitemOpen
  \bibfield  {author} {\bibinfo {author} {\bibfnamefont {T.}~\bibnamefont
  {Moriya}},\ }\bibfield  {title} {\enquote {\bibinfo {title} {{Anisotropic
  Superexchange Interaction and Weak Ferromagnetism}},}\ }\href {\doibase
  10.1103/PhysRev.120.91} {\bibfield  {journal} {\bibinfo  {journal} {Phys.
  Rev.}\ }\textbf {\bibinfo {volume} {120}},\ \bibinfo {pages} {91} (\bibinfo
  {year} {1960})}\BibitemShut {NoStop}%
\end{thebibliography}%

\end{document}